\documentclass[%
 reprint,
nofootinbib,
 amsmath,amssymb,
 aps,
]{revtex4-2}

\usepackage{bm}
\usepackage{latexsym}
\usepackage{dcolumn}
\usepackage{amsmath,amsfonts,amssymb}
\usepackage{graphicx,epsfig}
\usepackage{psfrag}
\usepackage{amsthm}
\usepackage{color}

\usepackage{nccmath}
\usepackage{moresize}
\usepackage{enumerate}
\usepackage[hang,flushmargin]{footmisc}
\usepackage[titletoc,toc]{appendix}
\usepackage{lipsum}
\usepackage{subcaption}
\usepackage{epstopdf} 
\usepackage{hyperref}
\usepackage{rotating}
\usepackage{multirow}
\usepackage{mathtools}

\usepackage{stackengine}
\usepackage{scalerel}

\newcommand{\be}{\begin{equation}}
\newcommand{\ee}{\end{equation}}
\newcommand{\bea}{\begin{eqnarray}}
\newcommand{\eea}{\end{eqnarray}}
\newcommand{\bse}{\begin{subequations}}
\newcommand{\ese}{\end{subequations}}
\newcommand{\bce}{\begin{center}}
\newcommand{\ece}{\end{center}}
\newcommand{\bfg}{\begin{figure}}
\newcommand{\efg}{\end{figure}}
\newcommand{\bit}{\begin{itemize}}
\newcommand{\eit}{\end{itemize}}
\newcommand{\bed}{\begin{description}}
\newcommand{\eed}{\end{description}}
\newcommand{\ben}{\begin{enumerate}}
\newcommand{\een}{\end{enumerate}}
\newcommand{\nn}{\nonumber}

\newcommand{\fr}{\frac}

\def\le {\left}
\def\ri {\right}
\def\a  {\alpha}
\def\b  {\beta}
\def\c  {\gamma}

\def\d  {\delta}

\def\k  {\kappa}

\def\m  {\mu}
\def\n  {\nu}

\def\p  {\pi}

\def\vth {\vartheta}

\newcommand{\Uself}{U_{\text{\scriptsize self}}}

\newcommand{\bdm}{\begin{displaymath}}
\newcommand{\edm}{\end{displaymath}}

\begin{document}


\title{A singularity free classical theory of gravity}
\author{Saurya Das} 
\email{saurya.das@uleth.ca}
\author{Mitja Fridman} 
\email{fridmanm@uleth.ca }
\affiliation{
Theoretical Physics Group, Department of Physics and Astronomy, University of Lethbridge, \\
4401 University Drive, Lethbridge, Alberta T1K 3M4, Canada
}%

\author{Sourav Sur}
\email{sourav@physics.du.ac.in}
\affiliation{
Department of Physics and Astrophysics, University of Delhi, \\
Delhi 110007, India
}%


\begin{abstract}
We present a classical theory of gravity, which is singularity free at short distances and reduces to General Relativity at large distances. We discuss its implications. 
\end{abstract}

\maketitle

\section{Introduction}

Despite the remarkable success of Newtonian gravity and its covariant counterpart, General Relativity (GR), in explaining large scale phenomena from the planetary to the cosmological scales, it is well-known that they both possess unavoidable singularities at short distances, where the theories break down. In Newtonian gravity, this can be seen from the expressions of gravitational potential 
of a point particle of mass $M$ at a distance $r$,
and the self-energy of a gravitationally bound sphere of radius $\varrho$, given respectively by 
\cite{kittelmechanics}
\bea
&& V(r) =\, -\, \fr{GM}{r} \,,
\label{potential1} \\
&& \Uself = - \frac{3 G M^2}{5\varrho} \,,
\label{self1}
\eea
where $G$ is the Newton's constant. One can see 
that $V \to \infty$ as $r\rightarrow 0$, and $\Uself \to \infty$, as 
$\varrho \to 0$, i.e. when the spherical gravitating object reduces to a point particle. 
Thus, for example, the gravitational self-energy of an electron is infinite in standard Newtonian gravity!

In the context of GR, the singularity can most easily be seen from the expression for the Kretschmann scalar for the
vacuum Schwarzchild solution of gravitating mass $M$,
expressed as the square of the Riemann curvature tensor
\be
K \equiv\, R_{\a\b\c\d} R^{\a\b\c\d} =\, \fr{48 \,(GM)^2}{c^4 r^6} \,,
\ee
%
which blows up as $r \to 0 $  
\cite{kretschmann}.
Furthermore, the celebrated singularity theorems of Penrose and Hawking, together with their essential ingredient, the Raychaudhuri equation, show that for globally hyperbolic space-times satisfying the weak energy condition, and having trapped surfaces, there exist incomplete geodesics. This has been accepted as the most definitive sign of a singular spacetime 
\cite{RE,singularity}.
For the Schwarzschild black hole, for e.g., the geodesics entering the horizon reach the singularity at $r=0$ in a finite proper time or affine parameter, and hence incomplete. 

While analogous singularities arise in electrodynamics and Yang-Mills theories, they are effectively resolved by their quantum counterparts,
thus enabling finite predictions for physical processes.
Quantum mechanically, potentials and fields remain finite, as is the electromagnetic self-energy of an electron, and there is no equivalent of the singularity theorems to which these theories are subjected 
\cite{peskin}.
Unfortunately for gravity, there is no such satisfactory renormalizable quantum theory yet, capable of making finite predictions, in spite of numerous attempts and progress being made in various approaches for over half a century
\cite{string,loop,kiefer,asympsafety,euclidean,hamber,causal}. 

In light of the above, it makes sense to revisit classical gravity to determine whether a singularity is indeed unavoidable 
in a theoretical formulation. 
As we shall show in this article, the answer is remarkably in the negative, and present a classical theory of gravity, which is covariant, and reduces to the standard GR at large distances while being devoid of singularities at all distances. We shall prove this by examining all the tests mentioned above for the existence of singularities.  

\section{Non-singular potentials: Newtonian Theory}

{We begin by proposing a simple modification of the Newtonian potential $V(r)$ in Eq.\,(\ref{potential1}), such that the following conditions hold
\bea
&& V(r) =\, - \fr{GM} r \,,   
~\text{as}~ r \to \infty \,,
\label{potential2a} \\
&& V(r) \to\, \text{constant},~~\text{as}~ r \to 0 \,,
\label{potential2b} \\
&& V^{(1)}(r), \, V^{(2)}(r),\, \dots,\, V^{(n-1)}(r) \to 0 \,,~~\text{as}~ r \to 0 \,,  \qquad
%
\label{potential2c} \\
%
%
&& |V(r)|,\,  |V^{(1)}(r)|,\, |V^{(2)}(r)|,\, \dots
< \infty~,~~ \forall r ~, 
\label{potential2e} 
\eea
where $V^{(n)}(r) \equiv d^n V(r)/dr^n$. 
It may be noted that the gravitational field $g(r) = - V^{(1)}(r)$ and the curvature scalar $R(r) \propto - V^{(2)}(r)$. 
}
Eqs.\,(\ref{potential2a})\,-\,(\ref{potential2e}) follow from the requirement
that the modified potential must reduce to the Newtonian potential 
at large distances, and must approach a constant at short distances. By `short' we mean a length scale of $\ell \simeq 10^{-4}$\,m or less, since the Newton's law of gravity has been well-tested 
till about $10^{-4}$\,m.
In addition, if the interaction of gravity with the 
standard model fields is considered, the upper bound on the length scale could be considerably smaller.
Furthermore, we require up to the $(n-1)^{\text{th}}$ derivative of the potential $V(r)$ to vanish at $r=0$, assuming that the corresponding equation of motion (Newtonian or relativistic) involves the $n^{\text{th}}$ derivative of $r$. We also require $V(r)$ and its derivatives to be bounded everywhere.

Eq.\,(\ref{potential2b}) automatically removes the short distance singularity associated with Eq.\,(\ref{potential1}). Furthermore, the self-energy of a gravitationally bound point particle of mass $M$ and density $\rho(r) = M\, \delta^{3}(r)$ 
is given by 
%
\be
\Uself = \int d^3r \, \rho(r) \, V(r) =\, M \,V(0) =\, \text{constant},
\ee
which clearly gets rid of the infinity in Eq.\,(\ref{self1}).
%
%
%
%

The underlying non-relativistic theory associated with a potential of this type is encapsulated in the Poisson equation for the modified
theory, which governs the field dynamics. This can be formally expressed as 
\be \label{poisson0}
[{\tilde V} (-i \vec \nabla)]^{-1} V (\vec r) = \bar\k \rho \,, 
\ee
where $\tilde V(k)$ is the Fourier transform of the modified potential $V(r)$ and $\bar\kappa$ is a constant. 
Its explicit form, for power law potentials is given in Appendix\,\ref{appendixa}. 

\section{Relativistic Theory: $f(R)$ Gravity}

Let us proceed to embed the above in a covariant theory. Note first that since in dealing effectively with weak gravity, as seen from the conditions (\ref{potential2a})\,-\,(\ref{potential2e}), one can identify as usual
\be \label{metricpot}
|g_{00}| =\, 1 +\, 2V \,, \quad (\text{setting}~ c = 1).
\ee 
Then from Eq.\,(\ref{potential2a}) it follows that at large distances one has the standard theory of GR. 

{
Next, to determine the correct theory at short distances, let us consider $f(R)$ gravity, where $f(R)$ is a suitable function of the curvature scalar $R$, so that acts the gravity action is 
\cite{frreview1,frreview2,multamaki,kalita}
\be
S  =\,  \fr 1 {16 \pi G} \int d^4 x \sqrt{- g}\, f(R) \,,
\ee
where $g$ denotes the metric determinant. Assuming spherical symmetry, we have the metric tensor of the form
\be \label{metric}
g_{\m\n} =\, \text{diag} \le(- s(r),\, p(r),\, r^2,\, r^2\sin^2\vth\ri) \,
\ee 
where $s(r)$ and $p(r)$ are certain functions of the radial variable $r$, that need to be determined by solving the field equations. 
As shown in the Appendix B, for weak-curvature space-times, or  equivalently 
for $|V(r)| \ll 1 \,$, 
one can write 
\be 
s(r) =\, 1 + 2 V(r) =\, 1 +\, \d s(r) \,,
\ee 
i.e., treat $2 V(r)$ as a perturbation $\d s(r)$ over the flat (Minkowski) solution of the gravitational field equations in GR (with $f(R) = R$). 
The following quantities then become
\bea  
&& X(r) \equiv\, s(r) \, p(r) =\, 1 +\, \d X(r) \,,\\
&& F(R) \equiv\, \fr{df}{dR} =\, 1 +\, \d F(R) \,, 
\eea 
with
\be \label{deltaF}
\d F(R(r)) =\, -\, \d s(r) +\, A  =\, -\, 2 V(r) +\, A \,,  
\ee
where $A$ is an integration constant which may suitably absorb any additive constant in the potential $V(r)$. As such, expressing $\, 
f(R(r)) =\, R(r) +\, \d f(r)$, one gets
\bea \label{df}
\d f(r) \!&=&\! \int dR(r) \,\d F(R(r)) \nn\\
\!&=&\! \int dr \le[- 2 V(r) + A\ri] \fr{dR}{dr} \,,
\eea 
by computing the curvature scalar $R(r)$ using the metric perturbations $\delta s(r)$ and $\delta X(r)$. If the function $R(r)$ can be inverted, i.e., one is able to express $r = r(R)$, then by substituting the same in Eq.\,(\ref{df}), the precise form of $\d f(R)$ can be obtained. 
}
To examine this further, we shall classify the function $V(r)$ into two mutually exclusive and exhaustive types in what follows.

\subsection{$V$ is an analytic function of $r$}

%
%

In this case, one can write, near $r=0$
\begin{eqnarray}
    V(r) &&= V(0) + V^{(1)} r + \frac{1}{2!}  V^{(2)} r^2
    + \frac{1}{3!}  V^{(3)} r^3 + \dots 
    %
\label{potentialtaylor}
\end{eqnarray}
%
%
Depending on the order of the first non-zero derivative, 
the above translates to $\delta s$ of the form
\begin{eqnarray}
    \delta s = ar^{q}~, \label{deltas} 
\end{eqnarray}
%
where $a$ is a constant.
For the above, as shown in the Appendix \ref{sphersymmfR}, one has
\begin{eqnarray}
    {R=-3q(q+1)ar^{q-2}}\,
\end{eqnarray}
{and}
\begin{eqnarray}
    %
    %
    f(R)  &&=  
%
     R  + 
{\bar c \,} R^{2(q-1)/(q-2)}
    \label{fR1} \\
    \bar c&&=
    \frac{q-2}{2(q-1)}[3q(q+1)]^{-q/(q-2)} |a|^{-2/(q-2)} 
    \label{fR1a}
\end{eqnarray}
%
%
Comparing Eqs.(\ref{deltas}) and (\ref{potn}), we see that $a<0$, resulting in the $|a|$ in Eq.(\ref{fR1a}) above.
Now, $q$ must be a positive integer, as otherwise
a certain derivative of the metric, along with all higher-order derivatives, would diverge as $r\rightarrow 0$.
This would cause the connection, the curvature, or some order derivative of these to blow up,
%
render an expansion of the form of 
Eq.(\ref{potentialtaylor}) invalid, taking it outside the domain of analytic functions.
%
%
%
Furthermore, if one also requires $\bar c$ to be positive, then  
one has $2 < q \leq \infty$. 
Several comments are in order. 
First, to ensure that the power of $R$ in the second term in Eq.(\ref{fR1})
is positive, such that the modified theory is local, we require  
$q>2$
%
\cite{ghost1,stelle}. Second, although a large range of $q$ is allowed, the value $q\rightarrow \infty$
seems special, as in this case, one obtains the simple form 
$\delta f \sim R^{2+\delta}$, where $\delta\ll 1$.
We will discuss this further later.
%
%
%
%
%
%
%
%
%
%

We remind the reader that the action
(\ref{fR1}) is valid for short distances, which is our region of interest.
The complete action, valid for large distances,
which must be obtained by eliminating $r$
between the exact expressions of $R(r)$ and  
$f(r)=\int F (dR/dr)dr$, can be more complicated. 
%
The Kretschmann scalar close the origin is given by
\begin{eqnarray}
    {K= (3q^2-2q+7)(aq)^2 r^{2(q-2)}}
    \rightarrow 0,~\text{as}~r\rightarrow 0~,
\end{eqnarray}
Furthermore, since the metric and its derivatives are all finite for all values or $r$, as anticipated, no curvature singularity is encountered anywhere.

\begin{figure}
\includegraphics[scale=0.65]{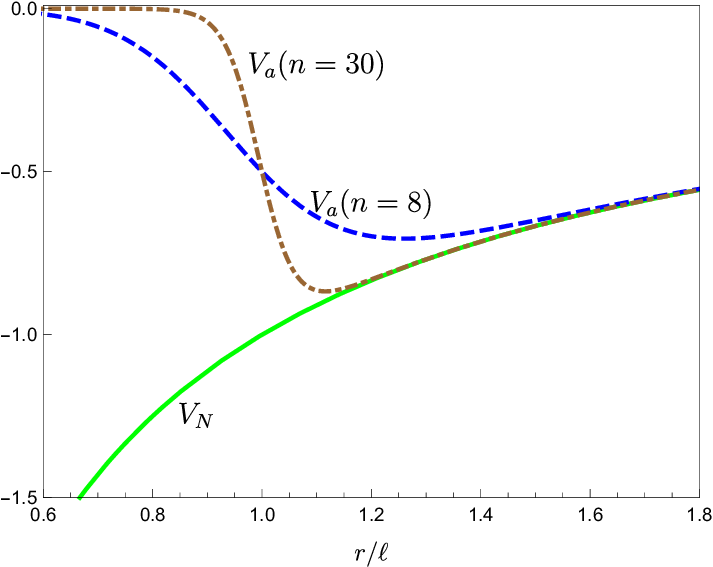}
\caption{\small 
Gravitational potential (in units of $GM/\ell)$ as a function of $r/\ell$, for two exemplifying analytic cases 
($V_a$ for $n = 8$ and $30$, shown by the dashed and dot-dashed lines respectively), and the Newtonian one ($V_N$, solid line). 
} 
\label{fig:pot6}
\end{figure}

\begin{figure}
\includegraphics[scale=0.65]{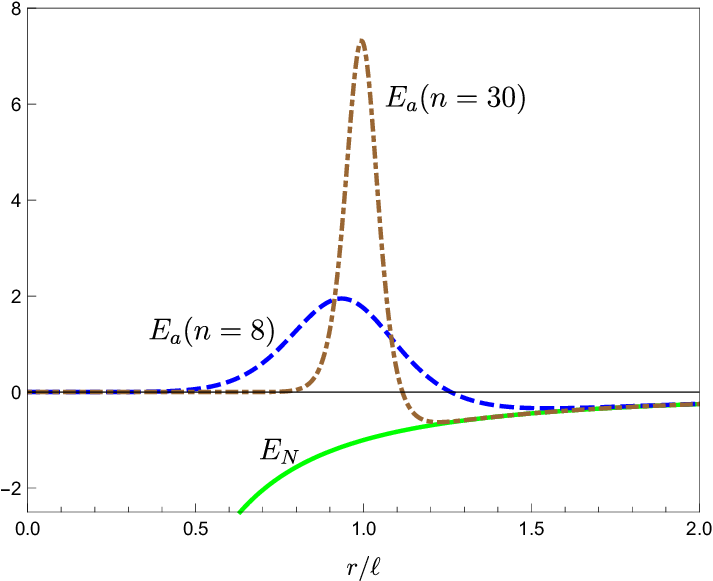}
\caption{\small 
Gravitational field (in units of $GM/\ell^2)$ as a function of $r/\ell$, for two exemplifying  analytic cases ($E_a$ for $n = 8$ and $30$, shown by the dashed and dot-dashed lines respectively), and the Newtonian one ($E_N$, solid line).
}
\label{fig:field6}
\end{figure}

A modified potential of the following type would fall into this class and satisfies all the conditions 
(\ref{potential2a}-\ref{potential2e})
\footnote{another class of modified potentials was considered in \cite{dassurmalady}}
\begin{eqnarray}
    V = -\frac{GM}{\ell}\,\frac{(r/\ell)^n}{1 + (r/\ell)^{n+1}}~,~n>0~.
    \label{potential3}
\end{eqnarray}
Eq.(\ref{potential3}) satisfies all the conditions (\ref{potential2a}-\ref{potential2e}),
and at large distances it obeys standard GR.
Near $r=0$ on the other hand, it can be expanded 
as 
%
\begin{eqnarray}
    V = \frac{GM}{\ell}
    \left[ 
    -\left(\frac{r}{\ell}\right)^n 
    + \left(\frac{r}{\ell}\right)^{2n+1} + \dots
     \right]~,
     \label{potn}
\end{eqnarray}
each of which terms, translated into the language of the metric via Eq.(\ref{metricpot}), 
clearly falls into the class (\ref{deltas}), starting with $q=n$.
Furthermore, it can be seen from Eq.(\ref{potn}) that 
the gravitational field $\vec g = -\nabla V >0$, signifying 
a repulsive force near $r \approx 0$. 
As can be seen from Figs.(\ref{fig:pot6}) and (\ref{fig:field6}), 
the maximum value of the short-distance repulsive field and the width of the `bump', are  
monotonically increasing and decreasing functions of $n$, respectively. In fact, it can be shown that for $n\gg1$, which we assume to be the case, 
the maximum field
is given by $g_{max} = (GM/\ell^2)(n/4+\mathcal{O}(1))$. Therefore, if this maximum can be estimated, for example in a laboratory experiment (for a given mass $M$), then with some knowledge of the length scale $\ell$, the 
value of $n$ can be inferred from it as
$n \simeq {4g_{max}\, \ell^2/GM}$. 

\begin{figure}
\includegraphics[scale=0.65]{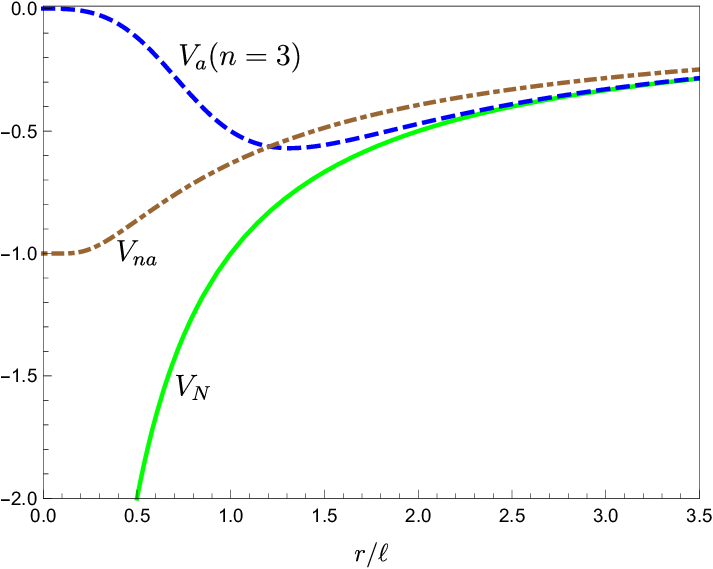}
\caption{\small 
Gravitational potential (in units of $GM/\ell)$ as a function of $r/\ell$, for the Newtonian case ($V_N$, solid line), non-analytic case ($V_{na}$, dot-dashed line) and the analytic case ($V_a$, dashed line, considering $n = 3$ as an example).}
\label{fig:gr1}
\end{figure}

\subsection{$V$ is not an analytic function of $r$}

%
%

\begin{figure}
\includegraphics[scale=0.6]{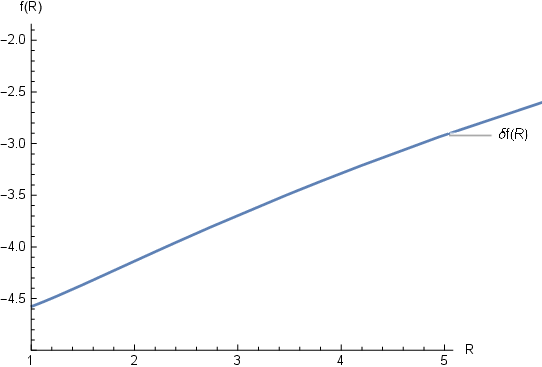}
\caption{\small 
The correction to the GR action, $\delta f(R)$, obtained by eliminating $r$ between 
Eqs.(\ref{Rnonanalytic}) and (\ref{deltafr}).
}
\label{fig:frparametric}
\end{figure}

In this case, a Taylor expansion of the form of
(\ref{potentialtaylor}) is not possible. We consider such a function 
\footnote{Metrics similar, but not identical to to that associated with potentials of this kind, along with the assumption of $|g_{tt}|=1/|g_{rr}|$, were considered in \cite{culetu1,culetu2,simpson}.}
\begin{eqnarray}
    V = 
    \frac{GM}{\ell} 
    \left( e^{-\ell/r} -1 \right)~,
    \label{potentialnonanalytic}
\end{eqnarray}
which satisfies all conditions
Eqs.(\ref{potential2a}-\ref{potential2e}). In subsequent calculations, the constant term in potential (\ref{potentialnonanalytic})
above is absorbed in the constant term in
$\delta F$ in Eq.(\ref{deltaF}). 
The expression for $R(r)$ for the current metric is given by
\be 
R =\, -\, \fr {2 G M \ell}{c^2} \le[1 + \le(1 - \fr{2 G M}{c^2 \ell}\ri)^{-1}\ri] \fr{e^{- \ell/r}}{r^4} \,,
\label{Rnonanalytic}
\ee 
%
%
%
The Kretschmann scalar for short distances is given by
\be 
K =\, \le(\fr{2 G M \ell}{c^2}\ri)^2 \le[1 + \le(1 - \fr{2 G M}{c^2 \ell}\ri)^{-2}\ri] \fr{e^{- 2 \ell/r}}{r^8} \,,
\ee 
which tends to vanish as $r \to 0$. In this case too, the metric and its derivatives are finite for all $r$
Hence, as before, no curvature singularity is encountered anywhere in spacetime. 
Furthermore, one has 
\begin{eqnarray}
    && \delta f(R(r)) = \int \delta F(R)\, dR \nonumber \\
    &&= \int \delta F(R)\, \frac{dR}{dr}\, dr \nonumber \\
    %
    %
  %
  && = \frac{(GM)^2}{\ell^2}
  \frac{e^{-2\ell/r}}{r^2}~.
  \label{deltafr}
\end{eqnarray}
%
%
{Eliminating $r$ between $R(r)$ in Eq.(\ref{Rnonanalytic}) and $\delta f(R)$ in Eq.(\ref{deltafr}) may be challenging to achieve analytically, as evident from the nature of those functions. However, it can be done numerically, as shown in our Fig.(\ref{fig:frparametric}), which shows that $\delta f(R)$ is a non-zero function, and that $f(R)$ differs from $f(R)=R$ for standard GR. }

\section{Lightcone Structure}

\begin{figure}
  \includegraphics[scale=0.32]{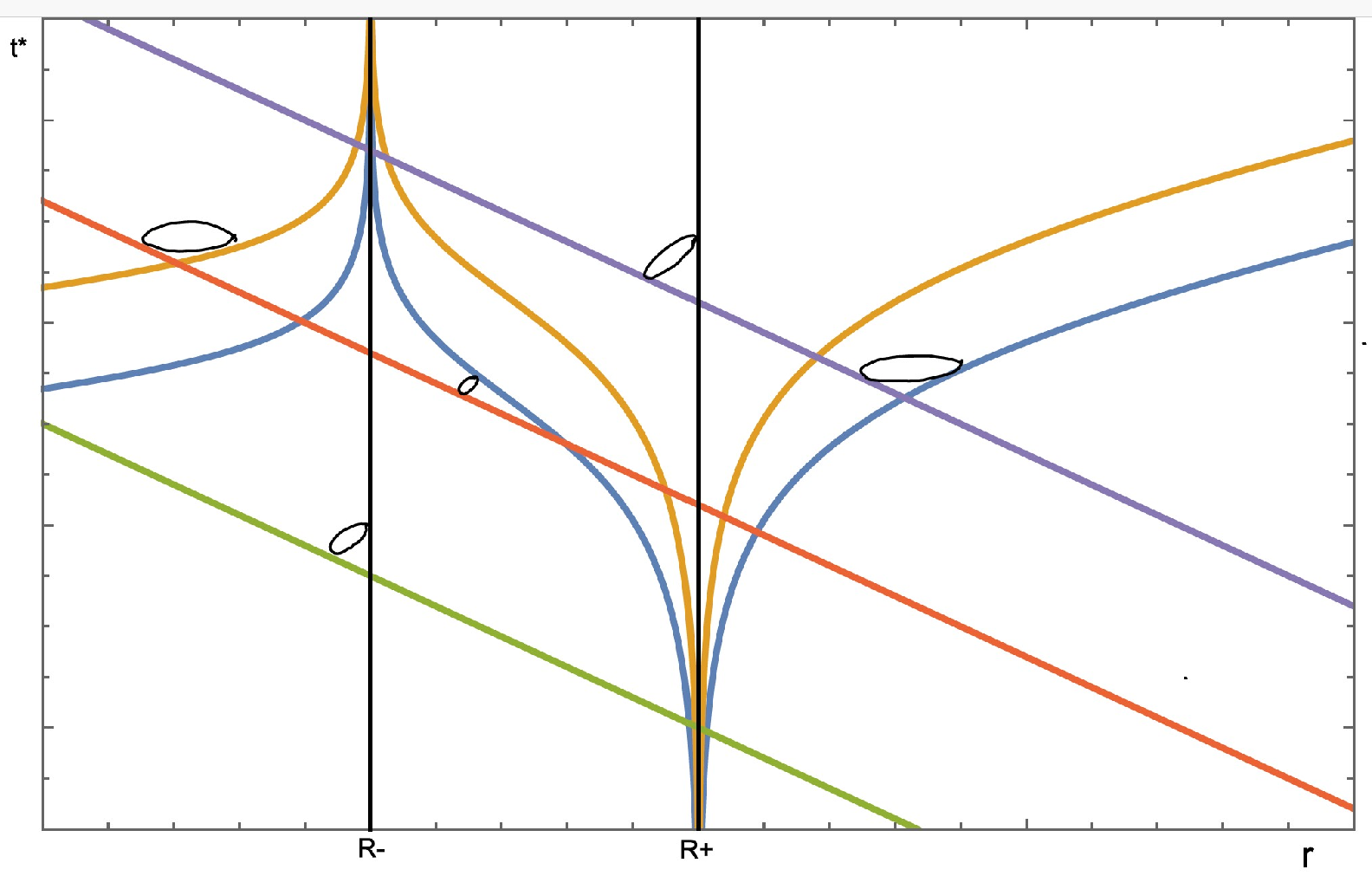}
  \caption{Light cone structure of spacetime defined by potential (\ref{potential3}). The straight lines are ingoing, and curved lines are outgoing null rays.}
  \label{fig:eddington}
\end{figure}

Finally, we look at the global lightcone structure of the spacetimes under consideration. 
For the analytical potential give in Eq.(\ref{potential3}), it can be easily shown that as long as $\ell$ is smaller compared to $GM$,
the equation $1+2V=0$ has exactly two real and positive solutions.
For 
simplicity, we choose $n=1$, although similar
results can be obtained for any $n$
\footnote{It was shown in ref.\cite{modesto} that {\it maximally extended} spacetimes defined by the potential  (\ref{potential3}) are geodesically complete for even $n$ (their $n$ and ours differ by unity.) 
. }
. 
For $n=1$, as long as $\ell <GM/2$,
the two solutions are given by
\begin{eqnarray}
    R_{\pm} = 
    R_0 \pm \sqrt{R_0^2 - 4\ell^2}~.
\end{eqnarray}
where $R_0=GM$. Note that when $\ell\rightarrow 0$,
$R_-\rightarrow 0$ and there is just the Schwarzschild horizon at $R_+=2R_0$.
Similar conclusions follow for $n>1$.
The corresponding tortoise, light cone and the time-coordinate $t^\star$ are given by
\begin{eqnarray}
&&    r_{\star} =\int \frac{dr}{1+2V}
\nonumber \\
&&=r    + \frac{1}{2 k_+} 
    \ln\left({{r-R_+}{}}\right)
    +\frac{1}{2k_-}\ln\left({{r-R_-}{}}\right)~ \\
&& u = t- r_\star \\
&& v = t + r_\star \\
&& t^\star = v - r = t + r_\star - r   
\end{eqnarray}
where 
$k_{\pm}= \frac{R_{\pm} - R_{\mp}}{2(R_{\pm}^2 + \ell^2)}$. 
Note that $k_+>0$ and $k_-<0$. 
The ingoing and outgoing geodesics are given by $v=t+r_\star=t^\star + r=$ constant and 
$u=t-r_\star=t^\star + r - 2r_\star =$ constant respectively.
In the Finkelstein, or $(t^\star,r)$, diagram, Fig.(\ref{fig:eddington}), 
the straight lines represent ingoing null-geodesics, while the curved lines, the outgoing null-geodesics. It can be seen from the light cone structure, that the outgoing null geodesics fall inwards inside the outer horizon 
$R_+$, signifying trapped surfaces, but they are no longer trapped inside the inner horizon. 
This bears resemblance to the causal structure of the Reissner-Nordstrom metric with two horizons, although unlike that case, here there is no singularity at $r=0$ for a geodesic to fall into, and disappear. Therefore, geodesics would continue to exist between $r=0$ and $r=R_-$ forever, and would not be incomplete. 

For the non-analytical potential given by 
Eq.(\ref{potentialnonanalytic}), there are no horizons, but there is no singularity either to hide them. In any case, there are no trapped surfaces and geodesics go on forever. 
We retain this example in our discussions because such a solution can in principle be realized in nature such as in an astrophysical scenario. In either case, we show that the spacetime in singularity free. 

Identical conclusions can be reached by considering the geodesic equation
\begin{eqnarray}
    \frac{d^2x^\mu}{d\tau^2} + \Gamma^{\mu}_{\nu\lambda} \frac{dx^\nu}{d\tau}
    \frac{dx^\lambda}{d\tau} = 0~. \label{geodesic}
\end{eqnarray}
It can be easily shown that for the 
metrics corresponding to potentials 
(\ref{potential3}) and (\ref{potentialnonanalytic}), the corresponding Christoffel symbols
vanish as $r\rightarrow 0$.
%
In other words, particles in this background become free as they approach $r=0$ and never encounter a singularity. 
%
%
This is reminiscent of asymptotic freedom in gauge theories, since here too a particle becomes effectively free from gravity as it approaches a massive particle \cite{afreedom}. 
However unlike in gauge theories, this  phenomenon is gravitational and purely classical in nature \cite{doug}. 
Other examples of singularity free
spherically symmetric solutions include
\cite{hayward,olmo1,olmo2,visser}.

\section{Conclusions}

To summarize, we have presented in this article a classical theory of gravity which reduces to GR at large distances, 
and remains singularity free everywhere. 
We demonstrated this by re-examining the gravitational field of a point particle at the origin, the self-energy of such a particle, the Kretschmann scalar for the corresponding spacetime at $r=0$,
and its causal structure.

One may have to revisit the question as to whether quantization of such a theory is compelling.
Furthermore, from Eqs.(\ref{fR1}) and (\ref{fR1a}), 
it follows that the coupling constant 
$G/\bar c$
in any perturbative quantum field theoretic expansion, has dimensions of $\text{(mass)}^{4/(q-2)}$, or equivalently 
$\text{(length)}^{-4/(q-2)}$ (in natural units). For example, for the potential (\ref{potential3}), for which $q=3$, it has 
the dimension of 
$
({\text{mass}})^4=\text{(length)}^{-4}$.
Similarly, in the limit $q\rightarrow \infty$, as is the case for $n\rightarrow \infty$,
it has the dimension of 
$
 {\text{(mass)}}^0$, or in other words an expected logarithmic dependence of the coupling constant with energy, as for gauge theories. 
This is to be contrasted with dimension of Newton's constant, namely
$
(\text{mass})^{-2}$, or $(\text{length})^2$, for which as one knows, the theory if perturbatively non-renormalizable.
Furthermore, for $q\rightarrow\infty$, one has $\delta f \sim R^2$, which corresponds to the Starobinsky model in the context of inflation \cite{star1,star2}, which has been argued to be a renormalizable theory \cite{starrenorm}.
However, since $q$ is expected to be large but finite, instead of an exact Starobinsky model, 
we anticipate the correct action to be
of the form $\delta f \sim R^{2+\delta}$, 
and correspondingly the coupling constant having dimensions 
$
(\text{mass})^{\delta}$, where $\delta\ll 1$.
In short, we see that the dimensions of 
the coupling constant of the theory 
seems to favour renormalizability of the theory \cite{asympsafety}. The details of these would need to be explored further. 

Finally, we note that since our modified potentials affect the Newtonian potential at small {\it and} at large length scales, we expect applications of our model to early and late time cosmology, as well as in astrophysics to yield new and potentially measurable results 
\footnote{See e.g. \cite{odintsov} for $f(R)$ cosmology.}. 
We hope to report on these elsewhere.

\begin{acknowledgments}
We thank V. Todorinov for discussions. 
We thank the anonymous referee for their valuable comments, which have helped improve the manuscript.
This work was supported by the Natural Sciences and Engineering
Research Council of Canada.
{
SS acknowledges financial support from Faculty Research Programme Grant -- IoE, University of Delhi (Ref.No./ IoE/ 2024-25/12/FRP).
}
\end{acknowledgments}

\appendix
 
\section{Poisson equation for modified gravitational potential}
\label{appendixa}
    
In this appendix, we find the Poisson equation for a modified gravitational potential of the form
%
\begin{eqnarray}
    V(r) = \epsilon\, e^{-\lambda r} r^p~,
    \label{vap1}
\end{eqnarray}
where $p \in \mathbb{R}$ and $\epsilon = \pm 1$, 
determined by  
whether the corresponding field is attractive or repulsive. 
Note that each term in the expansion (\ref{potn}) is of the above type.
$\lambda>0$ can be thought of as a regulator, with
$\lambda^{-1}$ giving the range of the force.
The Fourier transform of Eq.(\ref{vap1}) 
is given by 
\begin{eqnarray}
\tilde V(k) 
&&= \epsilon \int e^{-\lambda r} r^p
e^{-i\vec k\cdot\vec r}\,d^3r 
\nonumber \\
&& = \epsilon \int_0^{2\pi} 
\int_0^\pi 
\int_0^\infty 
e^{-\lambda r} r^p
e^{-ikr\cos\theta}\,r^2\sin\theta dr d\theta d\phi 
\nonumber \\
&& 
= \frac{2\pi \epsilon (p+1)!}{ik}
\left[  
\frac{-1}{(ik+\lambda)^{p+2}} 
+ \frac{1}{(-ik+\lambda)^{p+2}} \right]~.
\label{ft4}
\end{eqnarray}
The Poisson equation for any given potential 
can be written as
\begin{eqnarray}
[\tilde V(k)]^{-1} V (\vec r)= \bar\kappa \rho~, 
\label{poisson1}
\end{eqnarray}
where one substitutes 
$\vec k \rightarrow -i \vec\nabla$, 
and the constant $\kappa$ 
takes care of dimensional consistency of both sides. 
It can be easily verified that 
Eq.(\ref{poisson1}) reduces to the standard Poisson equation for the Newtonian potential, 
namely 
\begin{eqnarray}
 \nabla^2 V = 4\pi G\rho   
 \label{poisson2}
\end{eqnarray}
for $p=-1$, $\lambda=0$ and $\bar\kappa=G$.
Eq.(\ref{ft4}) can be simplified further via Taylor expansion for small $\lambda$
\begin{eqnarray}
\label{vks}
    && \tilde{V}(k)=-\frac{4\pi \epsilon}{(ik)^{p+3}}\,(p+1)!\,\,e^{-i\pi\frac{p+1}{2}} \cos{\!\left(\!\pi\,\frac{p+1}{2}\!\right)} \times \nonumber \\
    && \left[\vphantom{\cos{\!\left(\!\pi\,\frac{p+1}{2}\!\right)}}\,1+B_2(p)\!\left(\frac{\lambda}{ik}\right)^{\!\!2}\!+B_4(p)\!\left(\frac{\lambda}{ik}\right)^{\!\!4}\!+\cdots\right.  \\
    && \left.- i\tan{\!\left(\!\pi\,\frac{p+1}{2}\!\right)}\!\left\{B_1(p)\frac{\lambda}{ik}+B_3(p)\!\left(\frac{\lambda}{ik}\right)^{\!\!3}\!+\cdots\right\}\right]~, \nonumber
\end{eqnarray}
where $B_q(p)=\frac{(p+1+q)!}{q!(p+1)!}$. Note that in the case of $p\in\mathbb{Z}^+$, we are left only with even powers of $\lambda/ik$ for $p$ odd, and with odd powers of $\lambda/ik$ for $p$ even. Note that the above series is finite for $p=-1$ and $p=0$. The same cannot be said for other values of $p$.
Consequently, the inverse of the above reads as
\begin{eqnarray}
    &&[\tilde V(k)]^{-1} = -\frac{1}{4\pi\epsilon(p+1)!}\frac{e^{i\pi\frac{p+1}{2}}}{\cos{\left(\p\frac{p+1}{2}\right)}}\times \nonumber \\
    && \left[\vphantom{\tan{\!\left(\!\pi\frac{p+1}{2}\!\right)}}(ik)^{p+3}\!+K_2(p)\lambda^2(ik)^{p+1}\!+K_4(p)\lambda^4(ik)^{p-1}\!+\cdots \right. \\
    &&\left.+i\tan{\!\left(\!\pi\frac{p+1}{2}\!\right)}\!\left\{K_1(p)\lambda(ik)^{p+2}\!+K_3(p)\lambda^3(ik)^{p}\!+\cdots\right\}\right] \nonumber
\end{eqnarray}
which is valid for non-integer $p$ and odd integer $p$, where 
\begin{eqnarray}
    K_1(p)&=&B_1(p) \nonumber\\
    K_2(p)&=&-B_1^2(p)\tan^2{(\pi(p+1)/2)}-B_2(p) \nonumber \\
    K_3(p)&=&B_3(p)-2B_2(p)B_1(p)-B_1^3(p)\tan^2{(\pi(p+1)/2)} \nonumber \\ 
    K_4(p)&=&-B_4(p)+B_2^2(p)+B_1^4(p)\tan^4{(\pi(p+1)/2)} \nonumber \\
    &&+3B_2(p)B_1^2(p)\tan^2{(\pi(p+1)/2)}
\nonumber \\
&&-2B_3(p)B_1(p)\tan^2{(\pi(p+1)/2)}~,
\end{eqnarray} 
and so on. Note that the second line with the imaginary unit vanishes for odd integer $p$. For even integer $p$, the inverse of Eq. (\ref{vks}) reads as
\begin{eqnarray}
    &&[\tilde{V}(k)]^{-1}=\frac{1}{4\pi\epsilon(p+1)!\,i\lambda}\,\frac{e^{i\pi\frac{p+1}{2}}}{\sin{\left(\pi\frac{p+1}{2}\right)}B_1(p)} \times \\
    &&\left[(ik)^{p+4}-\frac{B_3(p)}{B_1(p)}{\lambda^2}{(ik)^{p+2}}-\left(\frac{B_5(p)}{B_1(p)}-\frac{B_3^2(p)}{B_1^2(p)}\right){\lambda^4}{(ik)^p}-\cdots\right]~. \nonumber
\end{eqnarray}
As can be seen, in general the Poisson equation that
it implies will be an infinite series and non-local (due to the inverse powers of $k$). Infinite derivative theories of gravity were encountered elsewhere \cite{anupam}. 

\section{Spherically symmetric vacuum solutions of $f(R)$ gravity}
\label{sphersymmfR}

We follow the notations of 
\cite{multamaki,kalita} and use the signature convention $(-,+,+,+)$. 
Given an unknown diagonal metric $g_{\mu\nu}$, the main $f(R)$ gravity equations are
\begin{eqnarray}
    {\cal L}_G &=& \frac{c^4}{16\pi G} f(R)~ \\
    F(R) &=& \frac{df(R)}{dR} \\
    g_{\mu\nu} &=& \text{diag} (-s(r),p(r),r^2,r^2\sin^2\theta)  \label{metric1} \\
    X(r)&=& p(r) s(r) \\
  2r\frac{F''}{F}&=&\left( \ln X\right)' 
 \left[ 2 + r \,\,\left( \ln F\right)' \right]   \label{freom1} \\
-4s +  4X &+& 2\left( \ln F\right)'(r^2 s'-2rs)
 + \left( \ln X\right)' (2rs - r^2 s')  \nonumber \\
 &+& 2r^2 s'' =0~.
\label{freom2}
\end{eqnarray}
Eq.(\ref{freom1}) can be rewritten as
\begin{eqnarray}
    \left( \ln X\right)'=
    \frac{2rF''/F}{2 + r \left( \ln F\right)'}~,
    \label{freom3}
\end{eqnarray}
and substituting it in Eq.(\ref{freom2}), we obtain
\begin{eqnarray}
 -2s +  2X  &+& \left( \ln F\right)'(r^2 s'-2rs)
 + \frac{r(2rs - r^2 s')F''/F}{2 + r \left( \ln F\right)'} 
 \nonumber \\
 &+& r^2 s'' =0~. \label{freom4}
\end{eqnarray}
Differential Eqs. (\ref{freom1}) and (\ref{freom2}) are independent of the metric signature, i.e., they remain as they appear above in both $(+,-,-,-)$ and $(-,+,+,+)$ signatures. Note that Eqs. (\ref{freom1}) and (\ref{freom2}) are equivalent to Eq. (14) of Ref. \cite{multamaki}, which is arduous to manipulate. Therefore, it is more convenient to use Eqs. (\ref{freom1}) and (\ref{freom2}), which are necessary and sufficient to satisfy the $f(R)$ modified Einstein equations.

As expected in the standard case, when $F=$ constant and $X=1$, 
Eq. (\ref{freom4}) reduces to
\begin{eqnarray}
r^2 s'' - 2s + 2 =0~~,
\end{eqnarray}
which has the solution 
\begin{eqnarray}
    s = 1 - \frac{k_1}{r} + k_2 r^2~.
    \label{scsoln}
\end{eqnarray}
The above shows just the Schwarszchild and the cosmological constant terms, which as we know is a solution for the standard Einstein equation with a cosmological constant, and where $f(R) =R$.

Since $V(r)$ vanishes in both $r \rightarrow 0$ and $r \rightarrow \infty$ limits
we can write the 
quantities $F,s$ and $X$ as perturbations over their standard values in Minkowski space, and as a solution of the Einstein action and Einstein equations of motion, as follows
\begin{eqnarray}
        F &=& F_0 + \delta F \label{Fperturb} \\
    \implies\,\ln F &=& \ln F_0 + \ln\left(1 + \frac{\delta F}{F_0} \right)  \nonumber \\
     &\simeq&  \ln F_0 + \frac{\delta F}{F_0} \\
    \implies\,(\ln F)' &\simeq& \frac{\delta F'}{F_0} \\ 
    s &=&1 + \delta s \\
    X &=& 1 + \delta X~, \\
    \implies\,(\ln X)' &\simeq& \delta X'~,
\end{eqnarray}
%
%
{where we assume $k_1=k_2=0$ so that $s$ in 
Eq.(\ref{scsoln}) is compatible with the potential 
(\ref{potential3}) at short and long distances respectively.}
With the above, Eqs. (\ref{freom3}) and (\ref{freom4}) assume the relatively simple forms namely
\begin{eqnarray}
 \delta X' &=&  \frac{r\,\delta F''}{F_0} \label{eom9} \\
\implies\,\,\delta X&=&\frac{1}{F_0}(r\delta F'-\delta F+A) \label{eom9.1} \\
-2 \delta s + 2 \delta X &-&\frac{2 r \delta F'}{F_0} + r^2 \frac{\delta F''}{F_0} + r^2\delta s''
=0 ~,\label{eom10}
\end{eqnarray}
where $A$ is a constant. 
Substituting Eq.(\ref{eom9.1}) in (\ref{eom10}), we obtain
\begin{eqnarray}
%
 &&-2 \delta s 
    -2\delta F + 2 A+ r^2 {\delta F''}{} + r^2\delta s''
=0~, \label{eom10.2}
\end{eqnarray}
which can be re-arranged as
\begin{eqnarray}
    r^2\delta F'' - 2\delta F 
= 2 (\delta s-A)  
 - r^2\delta s''~.
\label{eom12.3}
\end{eqnarray}
The above can be written as
\begin{eqnarray}
&& r^2[\delta F + (\delta s-A)]'' - 
{2[\delta F + (\delta s-A)]} =0 
\nonumber \\
 && i.e.,~r^2 Y'' - 2 Y =0  \nonumber
\end{eqnarray}
where $Y=\delta F + (\delta s -A)$.
%
Therefore, the solution for $Y$
(i.e., $\delta F$) is
\begin{eqnarray}
\label{dfgeneral}
    \delta F = -(\delta s- A) + \frac{c_1}{r} + c_2 r^2~.
\end{eqnarray}
In the following, we consider two solutions for the metric (i.e., $\delta s$), namely a power law solution and a non-analytical solution. They both reduce to the Minkowski space-time at $r=0$.

%
(a) Power law solutions for $\{\delta s, \delta X, \delta F\}$
exist. For example, if we assume 
$\delta s = a r^q$, then 
\begin{eqnarray}
    \delta F = -a r^q+\frac{c_1}{r}+c_2r^2+A~,
    \label{dfsolution}
\end{eqnarray} as implied by Eq. (\ref{dfgeneral}), and \begin{eqnarray}
    &&\delta X = (1-q)a r^q-2\frac{c_1}{r}
    +c_2r^2~,
\end{eqnarray}
as obtained from Eq. (\ref{eom9.1}).
%
%

The metric components and the Ricci scalar are then obtained as follows
\begin{eqnarray}
\delta p &=& \delta X-\delta s \nonumber \\
 &=& -qar^q-2\frac{c_1}{r}+c_2r^2 \\
 g_{00} &=& -(1+\delta s) = -1-ar^q \\
 g_{11} &=& 1 + \delta p= 1-qar^q-2\frac{c_1}{r}+c_2r^2 \\
 %
 %
 R&=&g^{\mu\nu}R_{\mu\nu}
 =-3q(q+1) ar^{q-2}
 ~,\label{R1}
\end{eqnarray}
where we obtained the Ricci tensor $R_{\mu\nu}$ by the standard definition. Note that in case of a signature change, the Ricci scalar $R$ changes sign.
From Eqs. (\ref{Fperturb}), (\ref{dfsolution}) and (\ref{R1}), and considering $c_1=c_2=A=0$, we obtain
\begin{eqnarray}
    F (R) &=& F_0-\Bar{F}_1 R^{q/(q-2)} \\
    %
    \implies\,f(R) &=& \int F(R)\, \mathrm{d}R \nonumber \\
    &=&F_0 R - F_1 R^{2(q-1)/(q-2)}~, 
\end{eqnarray}
where 
\begin{eqnarray}
    && \bar F_1 = a \left( \frac{1}{-3q(q+1)a}\right)^{q/(q-2)} \\
    %
&& F_1 = \frac{(q-2)}{2(q-1)} \bar F_1 
\end{eqnarray}
{As before, the constant $c_1=0$ follows from the perturbation assumption, since $\delta F$ must be small for small $r$ and $c_2=0$ follows from the fact that the cosmological constant term contributes only for large $r$. The constant $A$ adjusts the gravitational potential, and can be set to $A=0$ without loss of generality. Furthermore, as noted after Eq.(\ref{fR1a}), $a=-|a|<0$.  }

(b) Also, non-analytical solutions exist. For example $\delta s=2GM/c^2\ell 
    [\mathrm{exp}^{(-\ell/r)} -1]$, where we introduced a new short distance length scale $\ell$. By using the same procedure as above, we obtain the metric components as
\begin{eqnarray}
    \hspace{-0.5cm}g_{00}&=&-1-\frac{2GM}{c^2\ell}\left(e^{-\ell/r}-1\right) \\
    \hspace{-0.5cm}g_{11}&=&1-\frac{2GM}{c^2r}e^{-\ell/r} \\
    \hspace{-0.5cm}R&=&-\frac{2GM\ell}{c^2}\frac{\left(\left(1\!-\!\frac{2GM}{c^2\ell}\right)^{\!2}\!+\!\left(1\!-\!\frac{2GM}{c^2\ell}\right)\right)}{\left(1\!-\!\frac{2GM}{c^2\ell}\right)^{\!2}}\frac{e^{-\ell/r}}{r^4}~.
\end{eqnarray}
From the above we can see that obtaining $r(R)$ is non-trivial, which makes determining $\delta F$, and therefore $f(R)$, a challenge.


\end{document}